\documentclass[11pt]{article}
\usepackage{moriond,epsfig}

\def\ro{{\it ROSAT\/}}
\def\cha{{\it Chandra\/}}
\def\hst{{\it HST\/}}
\def\xmm{{\it XMM-Newton\/}}
\def\asc{{\it ASCA\/}}
\def\source{3EG~J1835+5918}

\def\be{\begin{equation}}
\def\ee{\end{equation}}
\def\bea{\begin{eqnarray}}
\def\eea{\end{eqnarray}}

\begin{document}
\vspace*{4cm}
\title{MULTIWAVELENGTH SEARCHES OF UNIDENTIFIED EGRET SOURCES}

\author{ N. MIRABAL$^{1}$, J. HALPERN$^{1}$, R. MUKHERJEE$^{2}$, E. 
GOTTHELF$^{1}$ \& F. CAMILO$^{1}$ }

\address{$^{1}$
Astronomy Department, Columbia University, 550 West 120th Street,
 New York, NY 10027}

\address{$^{2}$
Dept. of Physics \& Astronomy, Barnard College, Columbia University,
New York, NY 10027}

\maketitle\abstracts{3EG J1835+5918 
is the brightest of the so-called unidentified EGRET
sources at intermediate galactic latitude (l,b)=(89,25). We obtained
complete radio,
optical, and X-ray coverage of its error box, discovering a faint
ultrasoft X-ray
source in the ROSAT All-Sky Survey. Deep optical imaging at the location
of this
source, as pinpointed by an observation with the ROSAT HRI,
reveals a blank field to a limit of $V > 25.2$. The corresponding lower
limit on $f_{X}$/$f_{V}$ is 300, which signifies that the X-ray source 3EG
J1835+5918 is probably a thermally emitting neutron star. Here we report
on recent \cha\ and \hst\ observations that strengthen this identification.
3EG J1835+5918 may thus become the prototype of an hypothesized
population of older pulsars, born in the Gould belt, that can account
for as many as 40 local EGRET sources. In addition to 3EG 1835+5918, 
we review the
ongoing multiwalength effort by members of our group to study other
unidentified EGRET sources using X-ray, optical, and radio data.} 

\section{Introduction}

One important achievement of the EGRET mission was the discovery 
of 273 persistent high-energy ($> 100$ MeV)
$\gamma$-ray sources~\cite{ha}, of which 93 are
likely or possibly identified with blazars,
9 with rotation-powered pulsars, one with the radio galaxy 
Cen~A, and one with the LMC.
This leaves more than one half the
sources in the Third EGRET catalog 
without firm counterparts 
at other wavelengths. 
Many difficulties attend the identification of EGRET sources close to the
Galactic plane, but even at high Galactic latitude, the large error
circles and the lack of a tight relation between 
$\gamma$-ray flux and other properties such as X-ray flux and core radio
flux prevent all but the brightest counterparts from being identified 
securely on the basis of position alone.

Of the unidentified sources, approximately 
one third lie within
$|b| \leq 10^{\circ}$ along the Galactic plane.  This excess
of low-latitude sources must comprise a
Galactic population that is either similar to the already
identified $\gamma$-ray pulsars~\cite{ya}, 
or represents an entirely new class
of $\gamma$-ray emitters associated with the disk population.
The shapes of radio pulsar beams as determined by the
rotating vector model~\cite{ra}
demand that a fraction of young radio pulsars
are {\it not} visible from Earth.  The clear differences between
the broad observed $\gamma$-ray beam patterns and the narrow radio pulses
implies that $\gamma$-ray emission is probably visible from
a more complete range of directions than are the radio beams.
In addition to the Galactic plane population, it is speculated~\cite{hd}
that as many as 40 of the steady, unidentified EGRET sources located at
intermediate Galactic latitude are a population of older pulsars
associated with the Gould Belt, an inclined, expanding disk of star formation
in the solar neighborhood that is $\approx 3 \times 10^7$~yr old.
The identification of Geminga as the first radio quiet pulsar~\cite{bc}
provides what might be the
prototype for several of the remaining unidentified Galactic sources. 

In spite of several $\gamma$-ray identifications, the nature 
of the majority of EGRET source remains mysterious.
A multiwavelength survey using X-ray, optical, and radio data,
is possibly the best available approach to determine 
the nature of unidentified sources. This technique has been 
used recently to find likely
identifications for several EGRET sources, such as
3EG J2227+6122~\cite{h1}, 3EG 1835+5918~\cite{m1}, and
the COS-B field 2CG 075+00, which overlaps with 
two EGRET sources 3EG J2016+3657 and 3EG
J2021+3716~\cite{mu1}. 
The possible discovery of 
new counterparts and envisioned applications
of $\gamma$-ray pulsar properties~\cite{ha1} have motivated the search for
additional EGRET identifications. However, the absence of obvious 
counterparts also admits
the possibility that there is another
population with characteristics unlike that of the identified EGRET 
sources yet to be discovered. A number of alternative $\gamma$-ray emitters 
have appeared in literature~\cite{s1,r1,t1}.
In the 
long term the advent of 
the next generation of high-energy $\gamma$-ray missions
{\it INTEGRAL\/}, {\it AGILE \/} and 
{\it GLAST\/}  will be able to explore the latter possibility. 
But prior to the new missions we have decided to take
advantage of refined multiwavelength techniques
and attempt 
detailed searches for counterparts of unidentified EGRET sources.
Here we present recent results of multiwavelength work on 
two unidentified sources: 3EG J1835+5918 and  
3EG J1621+8203. 

\section{3EG J1835+5918: The Next Geminga}

3EG J1835+5918 is the brightest of the so-called unidentified EGRET sources
at intermediate Galactic latitude ($\ell,b)=(89^{\circ},25^{\circ}$),
and the one with the smallest error circle, $12^{\prime}$ radius
at 99\% confidence.  It shows no
evidence for long-term variability~\cite{re1}.  Its spectrum can
be fitted by a relatively flat power law of photon index
--1.7 from 70~MeV to 4~GeV, with a turndown above 4~GeV.
Such temporal and spectral behavior is more consistent with a
rotation-powered pulsar than a blazar, which is the other major
class of EGRET source.  However,
observations 
find no radio pulsar in this field to an upper limit of 1~mJy at  
770~MHz~\cite{ni1}.  
No radio-source candidate has been found 
for \source, either Galactic or extragalactic.

We performed an exhaustive search for a counterpart
of \source, including deep radio, X-ray, and optical surveys,
as well as optical spectroscopic classification 
of every active object within {\it and outside} its error ellipse~\cite{m1}.
In summary, we identified all but one of the {\it ROSAT\/}
and {\it ASCA\/} sources in this region
to a flux limit of approximately
$5 \times 10^{-14}$~erg~cm$^{-2}$~s$^{-1}$, which is $10^{-4}$
of the $\gamma$-ray flux.  None of the identified sources 
are plausible $\gamma$-ray
candidates. The only unidentified source is RX~J1836.2+5925 that has no
optical counterpart.
RX~J1836.2+5925 was the brightest X-ray source within the
EGRET error ellipse (see Figure~1), with a flux of $1.6 \times 
10^{-13}$~erg~cm$^{-2}$~s$^{-1}
$.
Most important is the fact that this source falls on a blank optical
field to a limit of $R > 24.5$ and $V > 25.2$
(Figure~2). Our upper limit on the optical flux from RX~J1836.2+5925 implies 
that
its ratio of X-ray-to-optical flux $f_X/f_V$ is greater than 300,
an extreme that is seen only among neutron stars.  

As part of our survey, we discovered in the \ro\ All-Sky
Survey 
that RX~J1836.2+5925 was first detected a decade ago as a 
soft X-ray source~\cite{m2}. These archival \ro\ PSPC
photons provide the only X-ray spectral information, since 
the pointed \ro\ observations were obtained 
with the HRI, and the source was
not detected in the {\it ASCA} images.
Although only 22 photons were detected by the PSPC, they {\it all\/} fall
at energies below 0.4~keV.  In summary we have detected a 
soft X-ray, 
radio-quiet source, lacking a supernova remnant and with a high X-ray 
to optical flux. 
This is exactly what one would expect for
thermal emission from the surface of an older neutron star.
If fitted by a blackbody model,
such a pulse-height distribution is consistent with
$T \leq 5 \times 10^5$~K, but it is also dependent
upon the unknown intervening column density.  If we assume
$1 \times 10^{20} < N_{\rm H} < 3 \times 10^{20}$~cm$^{-2}$, the bolometric
flux corresponding to an assumed $T = 5 \times 10^5$~K is in the range
$(1.5-5.7) \times 10^{-13}$ erg~cm$^{-2}$~s$^{-1}$.  This is 10--40
times fainter than Geminga.
Thus it is either more distant than Geminga
($d > 160$~pc~\cite{bc}),
or cooler ($T < 5 \times 10^5$~K~\cite{hw}).
But a cooling neutron star of age $< 10^6$~yr should not be
{\it too} distant, because to place it $> 400$~pc above the Galactic
plane at $b = 25^{\circ}$
would require a kick velocity at birth of $> 500$~km~s$^{-1}$.
A reasonable upper limit on the distance is therefore 1~kpc,
which allows an isotropic $\gamma$-ray luminosity of
$6 \times 10^{34}\,(d/1\,{\rm kpc})^2$ erg~s$^{-1}$,
comparable to the spin-down power $I\Omega\dot \Omega$ of Geminga
($3.3 \times 10^{34}$ erg~s$^{-1}$). 
\begin{figure}
%\rule{5cm}{0.2mm}\hfill\rule{5cm}{0.2mm}
\vskip 0.0cm
\hskip 1.5truein
%\rule{5cm}{0.2mm}\hfill\rule{5cm}{0.2mm}
\psfig{figure=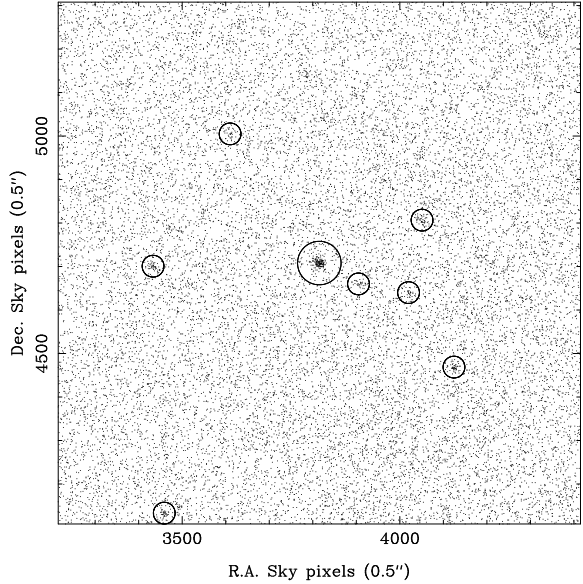,height=3.0truein}
\caption{
A $10^{\prime} \times 10^{\prime}$
portion of the $ROSAT$ HRI image centered on the location of the brightest,
unidentified X-ray source RX~J1836.2+5925({\it large circle}) in the
error circle of 3EG J1835+5918.
}
\end{figure}
\begin{figure}
%\rule{5cm}{0.2mm}\hfill\rule{5cm}{0.2mm}
\vskip 0.0cm
\hskip 1.5truein
%\rule{5cm}{0.2mm}\hfill\rule{5cm}{0.2mm}
\psfig{figure=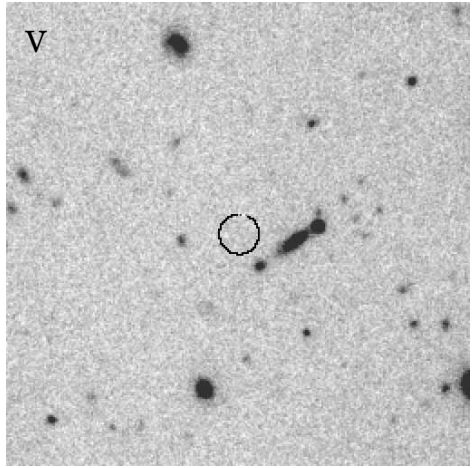,height=3.0truein}
\caption{
Zoom-in on a deep optical image from the MDM 2.4m
at the location of the unidentified
X-ray source RX~J1836.2+5925.  Seeing was $0^{\prime\prime}\!.8$,
and the 3$\sigma$ upper limit is $V > 25.2$.
The field is $70^{\prime\prime}$ across,
and the $ROSAT$ HRI error circle is
$3^{\prime\prime}$ in radius.  The X-ray astrometry was calibrated
using the optical positions of six well-identified X-ray sources
surrounding this field from Figure~1.
}
\end{figure}

Recently we have obtained coordinated observations of RX~J1836.2+5925
using the \cha\ ACIS 
and \hst\ STIS instruments~\cite{h2}
to strengthen its identification as a
neutron star by measuring its precise coordinates and using its 
X-ray spectrum and optical colors to address
the physics of the source. The preliminary results show that the
best-fitted X-ray spectrum requires two components, a soft 
thermal component previously 
suggested~\cite{m2} as well as 
a  hard tail that can be parametrized as a power law  which is the expected signature of a neutron star~\cite{h2}.
Deep optical imaging with \hst\ STIS reveals a blank field at the refined 
position of RX~J1836.2+5925 to 
a limit of $V>28.5$, corresponding to an extreme 
$f_{X}$/$f_{V}$ $> 6000$ ratio.  
In addition new radio observations using the Jodrell 
Bank 76 m Lovell telescope at a frequency of 1400 MHz 
find no radio pulsar to a flux density upper limit of $\approx$ 
0.1 mJy.
At a distance of 1 kpc, which represents a reasonable upper
limit for this source, the luminosity is then $\leq$
0.1 mJy kpc$^{2}$. For comparison Geminga has a luminosity 
upper limit of 0.06 mJy kpc$^{2}$ (cf. IAUC 5532). These limits are fairly 
sensible and would make  RX~J1836.2+5925 one of the faintest
radio pulsars if pulsations are ever found.
Summarizing, 
all the evidence thus far is consistent with an older or more distant 
version of the Geminga pulsar.
In the near future, it is important to look for X-ray pulsations 
from this object with \cha\ or \xmm\ to confirm it as a pulsar. 
A pulsar detection of  
RX~J1836.2+5925 might be feasible with a long observation using the
\cha\ High-Resolution
Camera if RX~J1836.2+5925 has a period  $P \geq 50$ ms.

The further identification of properties such as 
proper motion and distance should also serve
greatly to test the association of this EGRET source 
with the Gould Belt. In particular,
although some neutron stars might have been born in the Gould Belt, 
it is possible that
for high kick velocities the neutron star distribution is broader
than the Gould belt itself, as neutron stars travel to the outskirts 
of the Gould belt.

\section{3EG J1621+8203: An EGRET source coincident with NGC 6251}

3EG J1621+8203~\cite{mu2} was originally detected ~\cite{ha}
with a flux above 100 MeV of 
$1.1 \times 10^{-7}$~photon~cm$^{-2}$~s$^{-1}$. 
Its $\gamma$-ray spectrum can be fitted by
a power law of photon index --2.27 $\pm$ 0.53. In spite of uncertainties, 
such spectral index is slightly steeper than the 
typical EGRET $\gamma$-ray blazar. An 
examination of existing catalogs finds no 
radio-loud, flat-spectrum radio source within its
error box. However, no radio counterpart search was
carried out for this source~\cite{mh}, 
as it is in a region of sky not covered by the 5 GHz Green Bank survey.
3EG J1621+8203 is a rather ``weak'' MeV source with a large error 
ellipse of semi-major axis $a$=$1^{\circ}$  and 
semi-minor axis $b$=$0.7^{\circ}$ which makes the unique identification with a
counterpart a difficult task. An X-ray observation covering most of
the error ellipse of 3EG J1621+8203 was carried out with the 
 \ro\ PSPC instrument. Figure 3 shows the PSPC image
ovelaid
by the 1.4 GHz radio map of the same region. 
Complementary X-ray coverage of the periphery of the error ellipse
was obtained with additional pointings using the \ro\ HRI and 
\asc\ GIS instruments.  Several X-ray sources fall within the error
ellipse of 3EG J1621+8203. Using a multiwavelength approach   
we have been able to identify a number of active sources in the field  
with coronal emitting stars, radio-quiet 
QSOs and one with a galaxy cluster. The identifications are based 
mainly on a combination of catalog searches and optical imaging/spectroscopy 
of the X-ray positions.

An interesting source 
located within the error ellipse 
of  3EG J1621+8203 is the FR I radio galaxy NGC 6251. This source 
corresponds to the core and jet present in the 1400 MHz radio map (labels 
B1 through B5 in Figure 3). The highly extended jet 
has been observed at different scales and its 
inner structure has been resolved close to the core
using VLBI maps~\cite{jo}. NGC 6251 is an intriguing
object because of the possible link between BL Lac objects and 
FR I galaxies~\cite{ur}. 
It is believed that BL Lac objects correspond
to jets aligned with the line of sight. However it appears  
that the general properties of BL Lac objects 
are similar to FR I galaxies~\cite{ur}. Thus FR I galaxies 
could correspond to BL Lac with jets at non-zero viewing angles. 

In the 3EG catalog, Cen A (NGC 5128) is 
the only radio galaxy candidate not belonging to the EGRET blazar class
at energies above 100 MeV~\cite{se}. 
It is also the prototype of FR I galaxies and happens to be the 
brightest and nearest radio galaxy ($z=0.0018$, $\sim 3.5$ Mpc).
Cen A shows a jet that is oriented about  
 $70^\circ$ to our line of sight.
Its derived $\gamma$-ray 
luminosity is weaker by a factor of $10^{-5}$ compared to the typical EGRET
blazar. The low $\gamma$-ray luminosity and its radio properties 
might be evidence that Cen A is a misaligned BL Lac.

\begin{figure}
%\rule{5cm}{0.2mm}\hfill\rule{5cm}{0.2mm}
\vskip 0.0cm
\hskip 1.5truein
%\rule{5cm}{0.2mm}\hfill\rule{5cm}{0.2mm}
\psfig{figure=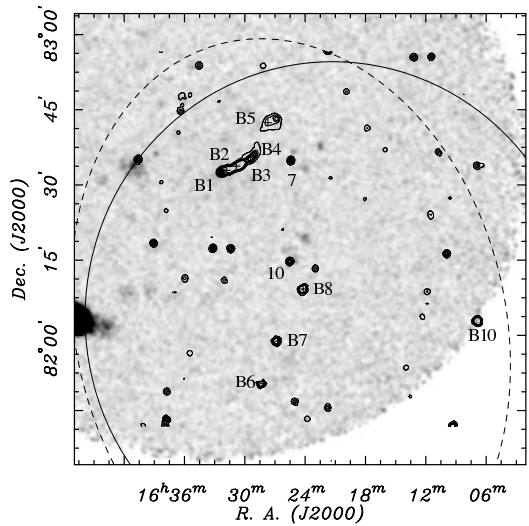,height=3.0truein}
\caption{
Contours of 1400 MHz sources in the field of 3EG J1621+8203
superimposed on the ROSAT PSPC image. 
}
\end{figure}

NGC 6251 shows some characteristics consistent with a BL Lac object. 
It has a flat-spectrum extended emission 
and inner radio core within uncertainties~\cite{sb}. Moreover
its radio luminosity is comparable to the average of EGRET blazar 
luminosity. Similar to Cen A, 3EG J1621+8203 is also less prominent in
$\gamma$-ray luminosity by a factor of $10^{2}-10^{5}$ than 
the typical EGRET blazar.
At an inclination angle
of 45 degrees to our line of sight~\cite{sd}, 
the luminosity of NGC 6251 whose jets are 
pointed away from our line-of-sight is expected to be less 
than the total amount of scattered energy, $F_1$, of an aligned jet.  
We can write the total scattered energy as a function of the viewing angle~\cite{we}:
 
$$F_1(s,\mu^*_s)= D^{3+s}(1-\mu_s^*)^{(s+1)/2},\eqno (1)$$
where the $\gamma$-ray flux seen by the observer is due to the scattered inverse
Compton emission of ambient low energy photons by highly relativistic 
particles in the jet. Particles are assumed to be electrons
and positrons distributed in energy as a power-law with a spectral index of
$s$,  $\mu_s^*$ is the cosine of the angle between the jet axis and the 
direction to the observer, and $D$ is the Doppler factor of the blob, defined 
as $D=\Gamma^{-1}(1-\beta\mu_s^*)^{-1}$, where $\beta c$ is the bulk velocity 
of the 
plasma. 

Figure 4 shows the decrease in scattered energy as a function of  
viewing angles, 
corresponding to two typical values of Lorentz factors ($\Gamma$) seen in 
blazars. This decrease could account for the discrepancy in $\gamma$-ray 
luminosities. Thus Cen A and NGC 6251 
could correspond to BL Lac objects with jets at 
different viewing-angles. The increased
sensitivity of future $\gamma$-ray missions may discover 
this new interesting class of objects associated with FR I galaxies
and provide a crucial observation to test unification models for AGN.

\begin{figure}
%\rule{5cm}{0.2mm}\hfill\rule{5cm}{0.2mm}
\vskip 0.0cm
\hskip 1.1truein
%\rule{5cm}{0.2mm}\hfill\rule{5cm}{0.2mm}
\psfig{figure=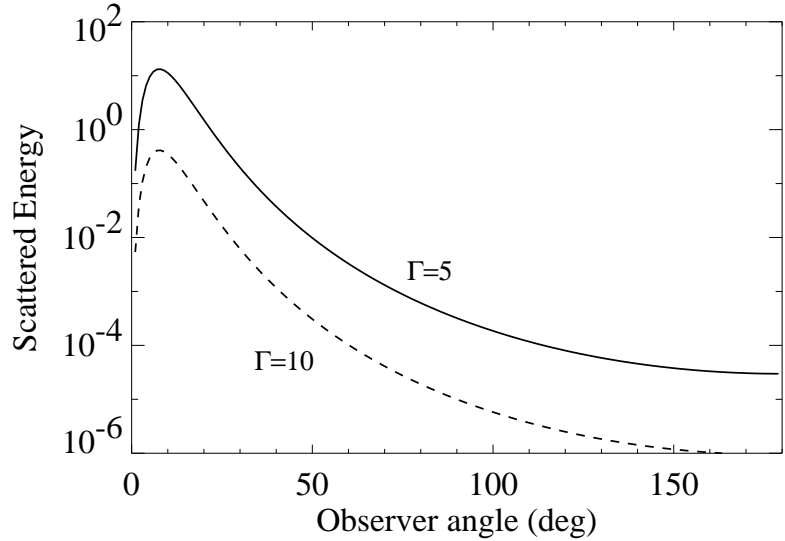,height=3.0truein}
\caption{
Decrease in the observed emission from a blazar as 
a 
function of jet orientation with respect to the observer.
}
\end{figure}

\section*{Acknowledgments}

\end{document}